\documentclass[12pt]{article}
\usepackage[top=25truemm,bottom=27truemm,left=22truemm,right=22truemm]{geometry}

\usepackage{indentfirst}
\usepackage[dvipdfmx]{graphicx}
\usepackage{float}
\usepackage{url}
\usepackage{siunitx}
\usepackage{array}
\usepackage{amsmath}
\usepackage{amssymb}
\usepackage{fancybox}
\usepackage{ascmac}
\usepackage{bm}
\usepackage{mathrsfs}
\usepackage{cancel}
\usepackage{physics}
\usepackage{color}
\usepackage{authblk}
\usepackage{appendix}
\usepackage{comment}
\usepackage{mhchem}
\usepackage[normalem]{ulem}
\usepackage{caption,latexsym,amsfonts,mathtools,here}
\usepackage{subcaption}
\usepackage{hyperref}
\usepackage{multirow}
\usepackage[
  backend=biber,
  sorting=none,
  style=numeric-comp,
  doi=false, url=true, eprint=true
]{biblatex}
\numberwithin{equation}{section}

\newcommand{\no}{\nonumber}

\newcommand{\hhh}{\hspace{6mm}}
\newcommand{\ds}{\displaystyle}

\newcommand{\eqns}{\hspace{-5pt}&=&\hspace{-5pt}}

\newcommand{\beqn}{\begin{eqnarray}}
\newcommand{\eeqn}{\end{eqnarray}}

\begin{document}
\begin{titlepage}
\begin{flushright}
{YITP-25-169, KOBE-COSMO-25-17}
\end{flushright}

\vspace{50pt}

\begin{center}

{\large{\textbf{Binary gravitational waves as probes of quantum  graviton states}}}

\vspace{25pt}

{Sugumi Kanno$^{1,2,3}$, Jiro Soda$^{4}$, and Akira Taniguchi$^1$}
\end{center}

\vspace{20pt}

\shortstack[l]
{\hspace{1.8cm}\it {\small $^1$Department of Physics, Kyushu University, Fukuoka 819-0395, Japan} \\[5pt]
\hspace{1.8cm}\it {\small $^2$Quantum and Spacetime Research Institute, Kyushu University}\\[5pt]
\hspace{1.8cm}\it {\small $^3$Center for Gravitational Physics and Quantum Information,} \\[2pt]
\hspace{2.1cm}\it {\small Yukawa Institute for Theoretical Physics, Kyoto University,} \\[5pt]
\hspace{1.8cm}\it {\small $^4$Department of Physics, Kobe University, Kobe 657-8501, Japan}}
\vspace{28pt}

\vspace{2cm}
\begin{abstract}

It is well known that the most reliable way to reveal the quantum nature of light is through photon number statistics, since photons exhibiting sub-Poissonian statistics unambiguously demonstrate their quantum behavior.
In this paper, we show that gravitons emitted by binary systems can, in principle, exhibit analogous sub-Poissonian statistics. 
The key idea is that the vacuum state of gravitons may not be the standard Minkowski vacuum but rather a nonclassical state imprinted with the physics of the early Universe, such as inflation. 
Accordingly, gravitational waves from binary systems provide a means to probe the graviton states generated in the early Universe.
As a concrete example, we show that squeezed graviton states originating from inflation can, in principle, imprint nonclassical graviton number statistics on gravitational waves from binary systems. In particular, we identify the frequency range in which the resulting coherent-squeezed graviton state can exhibit sub-Poissonian statistics. A realistic assessment of observational feasibility is left for future work.

\end{abstract}
\end{titlepage}

\tableofcontents

\section{Introduction}

One of the central challenges in fundamental physics is the detection of the graviton, which would provide direct evidence for the quantization of gravity.  
Dyson has argued that the enormous occupation number of typical gravitational waves makes the detection of a single graviton effectively impossible~\cite{Dyson:2013hbl}.  
This observation indicates that alternative approaches are needed to probe the quantum nature of gravity.

One possible approach is to exploit these large occupation numbers and search for gravitons indirectly through the quantum noise they induce in sensitive detectors~
\cite{Parikh:2020nrd,Kanno:2020usf,Parikh:2020kfh,Parikh:2020fhy,Kanno:2021gpt}. Alternatively, one may consider high-frequency gravitational waves, for which the occupation number is significantly smaller.  
In this regime, axion–magnon experiments have been proposed as potential detectors~\cite{Ito:2019wcb,Ito:2020wxi,Ito:2022rxn}, subsequently inspiring a variety of photon-based detection concepts~\cite{Ejlli:2019bqj,Berlin:2021txa,Domcke:2022rgu,Kanno:2023whr}. In addition, graviton–phonon conversion has been investigated as a potential mechanism for single-graviton detection~\cite{Tobar:2024bjr}.

The studies mentioned above primarily adopt the particle picture of gravitons.  
More recently, a graviton search based on photon–graviton quantum state conversion has been proposed~\cite{Ikeda:2025uae}, motivated by the finding that stimulated quantum state conversion can occur in a squeezed-state environment~\cite{Ikeda:2025qac}.
In this approach, the goal is to identify a graviton state that exhibits genuinely quantum behavior.

In this paper, following this line of thought, we propose a novel approach to graviton detection.  
Specifically, we suggest utilizing Hanbury Brown–Twiss (HBT) interferometry of gravitational waves from binary systems to probe the quantum nature of the graviton state.  
The basic idea is outlined below.

To begin with, it is worth noting that since the first detection of gravitational waves by LIGO in 2015~\cite{LIGOScientific:2016aoc}, all confirmed events have originated from binary systems.  
Gravitational waves from such systems are typically regarded as classical phenomena; however, they can also be described within a quantum-mechanical framework~\cite{Kanno:2025how}. In this description, the gravitational-wave state is naturally represented as a coherent state.  
The squeezing induced by binary systems has also been investigated, and, as expected, the effect is found to be very small.
Therefore, gravitational waves from a binary system are well described by a coherent state,
\begin{eqnarray}
   \ket{\psi_{\rm c}} = \exp[-i \int dt d^3 x  \left\{ T_{ij}(x^i, t)\,\hat{h}_{ij}(x^i,t)
    + \cdots \right\} ]  \ket{0_{\rm M}}\ ,
\end{eqnarray}
where $T_{ij}$ denotes the classical energy–momentum tensor of the binary system, and $\hat{h}_{ij}$ is the quantized metric perturbation operator describing the gravitational waves. The operator $\hat{h}_{ij}$ can be expanded in terms of graviton creation and annihilation operators, and its linear coupling to the classical source generates a coherent superposition of graviton states. The resulting quantum state $\ket{\psi_{\rm c}}$ thus represents the coherent graviton state associated with the classical source, while $\ket{0_{\rm M}}$ denotes the standard Minkowski vacuum.
The main observation of this paper is that 
the Minlowski vacuum $\ket{0_{\rm M}}$ may be replaced by a nontrivial quantum graviton state $\ket{Q} $ .
In that case, the gravitational-wave state generated by a binary system takes the form
\begin{eqnarray}
   \ket{\psi} = \exp[-i \int dt d^3 x  \left\{ T_{ij}(x^i, t) \hat{h}_{ij}(x^i,t)
    + \cdots \right\} ]  \big|  Q \rangle\ ,
\end{eqnarray}
The quantum graviton state $\ket{Q} $ encodes information about the early universe, so that the observed gravitational-wave state $\ket{\psi}$ inherits its quantum properties. For instance, phenomena associated with strong gravity — such as the generation of gravitational waves during a first-order phase transition — may imprint nontrivial features in the underlying quantum state $\ket{Q} $.
Although in this paper we focus on primordial gravitational waves generated during inflation, the present analysis is applicable to any mechanism that produces quantum states of gravitational waves~\cite{Kanno:2025how,Manikandan:2025dea,Guerreiro:2025sge,Dorlis:2025zzz,Dorlis:2025amf}.

As an illustrative example, we consider the squeezed state $\ket{\zeta}$ arising from inflation, where $\zeta$ denotes the squeezing parameter.
In this case, the quantum state of the gravitational waves emitted by a binary system can be modeled as a coherent–squeezed state,
\begin{eqnarray}
   \ket{\psi}  = \exp[-i \int dt d^3 x  \left\{ T_{ij}(x^i, t)\, \hat{h}_{ij}(x^i,t)
    + \cdots \right\} ]  \ket{\zeta}\ .
\end{eqnarray}
This construction captures both the classical radiation from the binary source and the quantum correlations originating from the primordial squeezed state.
In this paper, we demonstrate that the quantum nature of primordial gravitational waves can, in principle, be encoded in the graviton number statistics of gravitational waves emitted by binary systems. In particular, the graviton number statistics of the resulting coherent-squeezed state can become sub-Poissonian in an appropriate parameter regime, providing a clear indicator of nonclassicality. In an idealized setting, this information is captured by the second-order intensity correlation measured through Hanbury Brown-Twiss (HBT) interferometry. We emphasize, however, that the present work is conceptual and does not address whether such intensity-correlation measurements are feasible with current detectors. In particular, our analysis shows that it is possible to reveal the existence of a graviton state exhibiting quantum behavior.
The key point is that the graviton number statistics of the coherent–squeezed state can become sub-Poissonian in an appropriate parameter regime, providing a clear indication of the quantum nature of the state, as is well known in quantum optics~\cite{agarwal2013quantum}.
For a coherent field, the number statistics follow a Poisson distribution, for which the mean equals the variance, and consequently the Fano factor $F$ is unity, where the Fano factor $F$ is defined as the ratio of the variance to the mean of the particle-number distribution, 
\beqn 
F\equiv \frac{{\rm Var}(n)}{\langle n\rangle}\,.
\eeqn
It should be noted that the number statistics permitted by classical theories are always Poissonian or super-Poissonian, corresponding to a Fano factor $F\geq$ 1. 
Therefore, sub-Poissonian statistics ($F<1$) provide a clear signature of nonclassicality. To probe the graviton number statistics, one can employ intensity–intensity correlation measurements using Hanbury Brown–Twiss (HBT) interferometry~\cite{HanburyBrown:1956bqd,Brown:1956zza}.

We stress that the purpose of this paper is to identify a nonclassical target observable and the conditions under which it arises, rather than to provide a detector-ready proposal for current gravitational-wave observatories. In particular, the signature discussed below is not a modification of the mean strain waveform, but a higher-order statistical property of the radiation field, namely sub-Poissonian graviton number statistics.

The organization of the paper is as follows. In Section 2,we review the graviton number statistics in coherent–squeezed states and the Hanbury Brown–Twiss (HBT) interferometry, and outline a strategy for observing the nonclassical nature of primordial gravitational waves.
In Section 3,  we show that the quantum state of gravitons emitted from binary black holes can be described as a coherent–squeezed state. In Section 4, we estimate the frequency range in which the nonclassicality of primordial gravitational waves can be observed and discuss their detectability.
Section 5 is devoted to conclusions.
Throughout this work, we use natural units with $c=\hbar=1$.

\section{Hanbury Brown-Twiss interferometry}
\label{sec:coherent}
In this section, we review Hanbury Brown-Twiss ~(HBT) interferometry~\cite{HanburyBrown:1956bqd,Brown:1956zza}.
Hanbury Brown-Twiss interferometry was originally developed in the context of radio astronomy and used to measure stellar angular diameters with high precision by exploiting intensity-intensity correlations~\cite{HanburyBrown:1956bqd,Brown:1956zza}. In contemporary quantum optics, such correlations provide a powerful means of probing the nonclassical nature of light. The concept was first extended to consmology including relic gravitons~\cite{Giovannini:2010xg,Giovannini:2016esa,Giovannini:2017uty}, and more recently has been applied to the study of nonclassical primordial gravitational waves~\cite{Kanno:2018cuk,Kanno:2019gqw} and tests of quantum gravity theories~\cite{Kanno:2024gjt}. 
In what follows, we adapt the HBT formalism to the context of gravitational waves, focusing on how intensity correlations can reveal graviton number statistics.

\subsection{Fano factor}
As introduced in the previous section, the Fano factor $F$ serves as a useful indicator of nonclassicality in gravitational waves. 
For a coherent state, the graviton number distribution follows a Poisson distribution, for which $F=1$. In a classical theory, the fano factor is always Poissonian or super-Poissonian ($F\ge 1$). Therefore, a sub-Poissonian distribution ($F<1$) cannot be realized classically and serves as a reliable signature of nonclassicality. 

Let us consider the complex amplitude of an electromagnetic wave, denoted by $a$. The corresponding intensity is $I=|a|^2$. 
In classical theory, the normalized intensity-intensity correlation function is defined as
  \beqn
    g^{(2)}(\tau)=\frac{\langle I(t) I(t+\tau) \rangle}{\langle I(t)\rangle^2}\,.
    \eeqn
At zero time delay ($\tau=0$), the numerator reduces to $\langle I(t)^2 \rangle$. Using the relation for the intensity variance  $(\Delta I)^2 = \langle I(t)^2 \rangle - \langle I(t) \rangle^2 $,
and substituting this relation into the expression for $g^{(2)}(0)$, we obtain
\beqn
    g^{(2)}(0)=1+ \frac{(\Delta I)^2}{\langle I\rangle^2} \ .
    \eeqn
Since the variance $(\Delta I)^2$ is non-negative, we find 
\beqn
g^{(2)}(0)\ge 1,
\eeqn
with equality only when the intensity has no fluctuations.  
Thus, observing of $g^{(2)}(0) <1$ provides a clear indication of nonclassical behavior. In quantum theory, the corresponding correlation function is expressed in terms of field operators, allowing us to connect \(g^{(2)}(0)\) directly to the number statistics of quanta.

In quantum theory, the classical field amplitude $a$ is promoted to an operator $\hat{a}$ satisfying the commutation relation $[a,a^\dagger]=1$.
The intensity is represented by the expectation value of the number operator,
\beqn
I(t)=\langle a^\dag(t)a(t)\rangle\,. 
\eeqn
Consequently, the intensity-intensity correlation function, also referred to as the second-order coherence function, is defined as
    \beqn
    g^{(2)}(\tau)=
    \frac{\langle T: I(t) I(t+\tau)\!:\,\rangle}{\langle I(t)\rangle^2}
    =\frac{\langle a^\dag(t) a^\dag(t+\tau) a(t)a(t+\tau)\rangle}{\langle a^\dag(t)a(t)\rangle\langle a^\dag(t+\tau)a(t+\tau)\rangle}\,.
    \eeqn
Here, $T$ denotes time ordering and the colons represent normal ordering. The variable $\tau$ corresponds to the time delay between the signals received by two detectors.
The zero-delay coherence function $g^{(2)}(0)$ can be expressed in terms of the Fano factor as
    \beqn
    g^{(2)}(0)=1+\frac{(\Delta n)^2- \langle n \rangle}{\langle n \rangle^2}=1+\frac{F-1}{\langle n \rangle} \ ,
    \eeqn
where the Fano factor is defined $F=(\Delta n)^2/\langle n \rangle$.
Therefore, when the fano factor is less than unity, the zero-delay coherence function also satisfies $g^{(2)}(0)<1$. This implies that the detection of gravitational waves characterized by $F<1$, would indicate the nonclassical nature of primordial gravitational waves.
\subsection{Criterion for nonclassicality}
We now derive the condition under which the graviton number distribution of a squeezed–coherent state satisfies $F < 1$.  
It is well known that nonclassical behavior can be experimentally identified through Hanbury Brown–Twiss (HBT) interferometry.

The coherent state $\ket{\xi}$ is obtained by applying the displacement operator
\beqn
    \hat{D}(\xi)=
    \exp\left[\xi a^\dag -\xi^* a\right],\hhh \xi =|\xi| e^{i\theta} \ .
    \label{eq:opesqueezing}
    \eeqn
to the vacuum state, $\ket{\xi} = \hat{D}(\xi)\ket{0}$.
 The squeezed-coherent state $\ket{\psi}$ is defined as
 \beqn
    \label{eq:Bunchi}
    \ket{\psi} =\hat{S}(\zeta)\ket{\xi}\, 
    \eeqn
 where $\hat{S}(\zeta)$ is the two-mode squeezed operator,
 \beqn
    \hat{S}(\zeta)=
    \exp\left[\zeta^* a b-\zeta\,a^\dag b^\dag\right],\hhh \zeta =r e^{i\varphi} \ .
    \label{eq:opesqueezing}
    \eeqn

The expectation value of the graviton number operator is then
    \beqn
    \bra{\psi}n_a\ket{\psi}\eqns |\xi|^2\left[e^{-2r}\cos^2\left(\theta-\frac{\varphi}{2}\right)+e^{2r}\sin^2\left(\theta-\frac{\varphi}{2}\right)\right]+\sinh^2r\ ,
    \label{eq:expvalue_n}
    \eeqn
where we have use $\xi=|\xi|e^{i\theta}$. Assuming that the modes $n_a=a^\dag a$ and $n_b=b^\dag b$ are indistinguishable, we evaluate the standard deviation of their total occupation number. The variance is then given by
    \beqn\hspace{-5mm}
    (\Delta n)^2&=& 
    \langle 0|  \left( n_a + n_b  \right)^2  |0\rangle 
    -\langle 0|  n_a +n_b  |0\rangle^2 
    \nonumber\\
    &=&2|\xi|^2\left[e^{-4r}\cos^2\left(\theta-\frac{\varphi}{2}\right)+e^{4r}\sin^2\left(\theta-\frac{\varphi}{2}\right)\right]+4\sinh^2r+4\sinh^4r\,.
    \eeqn
For the phase alignment $\theta-\varphi/2=0$, the fano factor becomes
    \beqn
    F\eqns \frac{(\Delta n)^2}{\bra{\psi}n_a\ket{\psi}+\bra{\psi}n_b\ket{\psi}}=\frac{|\xi|^2 e^{-4r}+2\sinh^4r_k+2\sinh^2r}{|\xi|^2 e^{-2r}+\sinh^2r}\,.
    \eeqn
Therefore, the condition for the squeezed-coherent state to exhibit sub-Poissonian statistics is 
    \beqn
    |\xi|^2 \left(e^{-2r}-e^{-4r}\right) > \sinh^2 r + 2\sinh^4 r
    \label{eq:subPoisson_condition}
    \eeqn
In the case of primordial gravitational waves on super-horizon scales, where $r \gg 1$, the sub-Poissonian condition (\ref{eq:subPoisson_condition}) reduces to
    \beqn
    \label{eq:conditionsubPoisson}
    |\xi|^2>\frac{e^{6r}}{8}\ .
    \eeqn

The relation between the squeezed–coherent state and the coherent–squeezed state can be written as
    \beqn
    \label{eq:suqcoh}
    \ket{\psi} \equiv \hat{D}(\bar{\xi})\hat{S}(\zeta)\ket{0} 
    = \hat{S}(\zeta) \hat{D}(\xi) \ket{0} \ ,
    \eeqn
where the two displacement parameters are related by
\beqn
\xi = \bar{\xi} \cosh r + \bar{\xi}^*\,e^{i\varphi} \sinh r .
\label{eq:displacement_relation}
\eeqn

The condition for sub-Poissonian statistics in the coherent–squeezed ordering then becomes
\beqn
    |\bar{\xi}|^2>\frac{e^{4r}}{8}\ .
    \label{eq:condition_cohsq}
    \eeqn
Compared to Eq.~\eqref{eq:conditionsubPoisson}, the required coherent amplitude is exponentially smaller in the coherent–squeezed ordering, reflecting the enhanced quantum fluctuations induced by the squeezing operator.
\section{Coherent squeezed state of gravitons}
\label{sec:sec2}

This section bridges the quantum–cosmological origin of primordial gravitational waves with the quantum description of waves emitted by astrophysical binaries. 
We show that gravitational waves from binary black holes can be described as a coherent–squeezed state. 
This follows from the fact that primordial gravitational waves generated during inflation are in a squeezed state~\cite{Grishchuk:1989ss,Grishchuk:1990bj}, while binary black holes act as a source of coherent displacement of this state. 
As a result, the final quantum state of the emitted gravitational waves is coherent–squeezed~\cite{Kanno:2025how}.

\subsection{Squeezed state from Inflation}
\label{sec:inflation}
 Inflationary cosmology asserts that the large-scale structure of the Universe has a quantum origin.  Primordial gravitational waves are likewise generated from quantum fluctuations of spacetime. Consequently, the direct detection of primordial gravitational waves is one of the central goals in gravitational physics~\cite{Kawamura:2011zz, LISACosmologyWorkingGroup:2022jok}. 
The initial state of quantum fluctuations in the early universe is assumed to be the Bunch-Davies vacuum. In this subsection, we show that the Bunch-Davies vacuum appears as a squeezed state to an observer in the radiation-dominated era. This result provides the physical origin of the squeezing parameter $r_k$, which quantifies the quantum correlations imprinted on gravitational waves during inflation.

The Einstein–Hilbert action is given by
    \beqn
    \label{eq:EHaction}
	S_{\rm EH}=\frac{M_{\rm p}^2}{2}\int d^4x\sqrt{-g}R\ ,
	\eeqn
where $M_{\rm p}$ denotes the Planck mass and $g$ is the  determinant of the metric tensor $g_{\mu\nu}$.
At the linearized level, gravitational waves are described by the perturbed metric
    \beqn
    \label{eq:metric_a}
    ds^2=a^2(\eta)\left[-d\eta^2+(\delta_{ij}+h_{ij})dx^i dx^j\right]\ ,
    \eeqn
where $\delta_{ij}$ is the Kronecker delta,  $h_{ij}$ is the tensor perturbation, and $\eta$ denotes conformal time. The spatial indices $(i,j)$ run from 1 to 3, corresponding to the spatial coordinates $(x, y, z)$. The tensor perturbation $h_{ij}$ satisfies the transverse-traceless (TT) conditions $h^i{}_i=h^{i}{}_{j,i}=0$. Suppose that the transition from the inflationary epoch to the radiation-dominated era occurs at conformal time  $\eta=\eta_1$, the scale factor $a(\eta)$ in each era can be written as
    \beqn
    a(\eta)=
    \left\{ \,
    \begin{array}{lll}
    -\ds\frac{1}{H(\eta-2\eta_1)} & (-\infty <\eta<\eta_1)  & \hhh {\rm (Inflation)} \\[12pt]
    -\ds\frac{\eta}{H\eta_1^2} & (\eta_1<\eta) & \hhh {\rm (Radiation)}\label{eq:scalefactor}
    \end{array}
\right.
    \eeqn
Substituting the metric (\ref{eq:metric_a}) into the Einstein–Hilbert action (\ref{eq:EHaction}), we obtain the quadratic action for tensor perturbations,
    \beqn
    \label{eq:quadra_action_a}
	S_{\rm EH}\eqns\frac{M_{\rm p}^2}{8}\int d^4 x~a(\eta)^2 \left(h'_{ij}{h^{ij}}'-h_{ij,k}h^{ij,k}\right)\ ,
	\eeqn
where the prime denotes the differenciation with respect to the conforaml time $\eta$. Expressing $h_{ij}$ in terms of Fourier modes allows us to quantize each polarization mode as an independent harmonic oscillator.
The tensor perturbation $h_{ij}$ can be expanded in Fourier modes as
    \beqn
    \label{eq:hexpa}
    a(\eta)h_{ij}(\eta,\bm{x})=\frac{2}{M_{\rm p}}\frac{1}{\sqrt{V}}\sum_{\bm{k}}\sum_{P=+,\times}h_{\bm k}^P(\eta)e^{i\bm{k}\cdot\bm{x}}e_{ij}^P(\bm{k})\,,
    \eeqn
where $e^{P}_{ij}(\bm{k})\,,(P=+,\times)$ are the polarization tensors, normalized as
$e^{P}_{ij}(\bm{k})e^{Q}_{ij}(\bm{k})=\delta^{P Q}$.
The anihilation and creation operators obey the standard commutation relation. The wavevector $\bm{k}$ is discretized as $\bm{k}=(2\pi n_x/L_x, 2\pi n_y/L_y, 2\pi n_z/L_z)$, where $n_x, n_y, n_z$ are integers, $L_x, L_y , L_z$ are the box lengths, and the three-dimensional volume is $V=L_xL_yL_z$. The mode operator $h_{\bm k}^P(\eta)$ satisfies the equation of motion
    \beqn
    h_{\bm k}^{''P}(\eta)+\left(k^2-\frac{a''}{a}\right)h_{\bm k}^P(\eta)=0\ ,
    \eeqn
and can be expanded in terms of the creation and annihilation operators in each epoch as
    \beqn
    h_{\bm k}^P(\eta)=
    \left\{ \,
    \begin{array}{lll}
    b_{\bm k}^P v_k^{\rm I}(\eta) + b_{-\bm k}^{P\dag} v_k^{\rm I *}(\eta) & (-\infty <\eta<\eta_1)  & \hhh {\rm (Inflation)} \\[6pt]
    a_{\bm k}^P v_k^{\rm R}(\eta) + a_{-\bm k}^{P\dag} v_k^{\rm R *}(\eta) & (\eta_1<\eta) & \hhh {\rm (Radiation)}
    \label{eq:modeexpa}
    \end{array}
\right.\ ,
    \eeqn
where the mode functions are given by
    \beqn
    v_k^{\rm I}(\eta)\eqns\frac{1}{\sqrt{2k}}\left(1-\frac{i}{k(\eta-2\eta_1)}\right)e^{-ik(\eta-2\eta_1)} \ ,\\[6pt]
    v_k^{\rm R}(\eta)\eqns \frac{1}{\sqrt{2k}}e^{-ik\eta} \ .
    \eeqn
The transition from inflation to the radiation-dominated era leads to a mixing between positive- and negative-frequency modes, characterized by the Bogoliubov coefficients $\alpha_k$ and $\beta_k$. 
The vacuum state in each epoch is defined by $b_{\bm k}\ket{0}_{\rm I}=0$ and $a_{\bm k}\ket{0}_{\rm R}=0$, where $\ket{0}_{\rm I}$ is the Bunch-Davies vacuum. These two vacua are related through the Bogoliubov transformation,
    \beqn
    b_{\bm{k}}^P=\alpha_k^* a_{\bm{k}}^P-\beta_k^* a_{-\bm{k}}^{P\dag}\,,
    \eeqn
where the Bogoliubov coeﬃcients are given by
    \beqn
    \label{eq:Bogoliubov}
    \alpha_k= 1-\frac{1}{2k^2\eta_1^2}-\frac{i}{k\eta_1},\hhh \beta_k= \frac{1}{2k^2\eta_1^2}\,e^{2ik\eta_1}\,.
    \eeqn
These coefficients satisfy the normalization condition $|\alpha_k|^2-|\beta_k|^2=1$. It is often convenient to parameterize them as $\alpha_k=\cosh r_k$, $\beta_k=e^{i\varphi}\sinh r_k$, where $r_k$ is the squeezing parameter that quantifies the degree of squeezing of the vacuum state. The squeezing amplitude depends on the wavenumber as
  \beqn
    \label{eq:Bogoliubov}
    \sinh r_k= \frac{1}{2k^2\eta_1^2}\ .
    \eeqn
Using the Bogoliubov relation, we can connect the vacua in the two eras.
By applying the annihilation operator $b_{\bm{k}}^P$ to the Bunch-Davies vacuum $\ket{0}_{\rm I}$ and using the commutation relation $[a_{\bm{k}}^P, a_{\bm{k}'}^{Q\dag}]=\delta^{PQ}\delta_{\bm{k}, \bm{k}'}$, 
we obtain the relation between the two vacua $\ket{0}_{\rm I}$ and $\ket{0}_{\rm R}$:
    \beqn
    \label{eq:Bunchi}
    \ket{0}_{\rm I}=\hat{S}(\zeta)\ket{0}_{\rm R}\, 
    \eeqn
where the unitary squeezing operator is defined by
    \beqn
    \hat{S}(\zeta)=\prod_{\bm{k}}
    \exp\left[\zeta^* a_{\bm{k}}a_{-\bm{k}}-\zeta a^\dag_{-\bm{k}} a^\dag_{\bm{k}}\right],\hhh \zeta_k=r_k e^{i\varphi}\,.
    \label{eq:opesqueezing}
    \eeqn
Hence, the Bunch-Davies vacuum appears as a squeezed state in the subsequent radiation era.
Applying this operator to $\ket{0}_{\rm R}$, we can express the Bunch-Davies vacuum explicitly as
    \beqn
   \ket{0}_{\rm I}= \hat{S}(\zeta)\ket{0}_{\rm R}=\prod_{\bm{k}}\sum_{n=0}^\infty e^{in\varphi}\frac{\tanh^nr_k}{\cosh r_k}\ket{n_{\bm{k}}}_{\rm R}\otimes\ket{n_{-\bm{k}}}_{\rm R}\ .
    \eeqn
where $\ket{0}_{\rm R}=\ket{0_{\bm{k}}}_{\rm R}\otimes\ket{0_{-\bm{k}}}_{\rm R}$ and $\ket{n_{\bm{k}}}=\frac{1}{\sqrt{n!}}(a_{\bm{k}}^\dag)^n \ket{0_{\bm{k}}}_{\rm R}$. 
For large $r_k$, the Bunch-Davies vacuum becomes a highly entangled two-mode state. This two-mode entanglement between opposite momenta ($\bm{k}$, $-\bm{k}$) encodes the quantum origin of primordial gravitational waves, which later manifest as macroscopic stochastic backgrounds. Observational constraints on the squeezing parameter have been discussed in~\cite{Hertzberg:2021rbl}.

In the present analysis, we do not include the effects of environmental decoherence on primordial gravitational waves. Our purpose here is to study, at a proof-of-principle level, the graviton number statistics implied by a primordial squeezed state, assuming that sufficient quantum coherence survives. A quantitative assessment of how post-inflationary interactions and environmental decoherence affect the survival of squeezing and the resulting graviton number statistics is beyond the scope of the present paper and is left for future work.

\subsection{Coherent state from binary black holes}
We now turn to the astrophysical source of gravitational waves. In contrast to the primordial case, binary black holes generate classical gravitational radiation, which corresponds to a coherent quantum state of gravitons. We consider a binary black hole system following~\cite{Kanno:2025how}. The binary consists of two components with masses $m_1$ and $m_2$, whose worldlines are denoted by $\zeta_1$ and $\zeta_2$, respectively. For simplicity, we assume that the orbital trajectories $\bar{\bm{x}}_1(t)$ and $\bar{\bm{x}}_N(t)$ are given and neglect the backreaction due to gravitational-wave emission. The interaction between the  corresponding energy-momentum tensor and the graviton field gives rise to a coherent state of gravitons, which represents the classical gravitational waves emitted by the binary system.

Since the scale factor is unity ($a(\eta)=1$) for a binary black hole system, Eq.~(\ref{eq:metric_a}) reduces the metric to
    \beqn
	-d\tau^2\eqns ds^2=-dt^2 + \left(\delta_{ij}+h_{ij}\right)dx^idx^j\ ,\label{eq:metric}
	\eeqn
and, using Eq.~(\ref{eq:hexpa}) and (\ref{eq:modeexpa}), the tensor perturbation can be expanded as
    \beqn
	h_{ij}(t,\bm{x})=\frac{2}{M_{\rm p}}\frac{1}{\sqrt{V}}\sum_{\bm{k}}\sum_{P=+,\times}\left[\frac{e^{-i\omega_{\bm k} t}}{\sqrt{2\omega_{\bm{k}}}}e^{P}_{ij}(\bm{k})a^{P}_{\bm{k}}+\frac{e^{i\omega t}}{\sqrt{2\omega_k}}e^{P}_{ij}(-\bm{k})a^{P\dag}_{-\bm{k}}\right]e^{i\bm{k}\cdot\bm{x}}\,,
	\eeqn

The total action is obtained by adding the geodesic actions of the two point particles with masses $m_1$ and $m_2$ to the Einstein–Hilbert action~(\ref{eq:EHaction}):
    \beqn
    \label{eq:action}
	S\eqns S_{\rm EH}+S_{1}+S_{2}=\frac{M_{\rm p}^2}{2}\int d^4x\sqrt{-g}R-m_1\int_{\zeta_1} d\tau-m_2\int_{\zeta_2} d\tau\ ,
	\eeqn
where $\tau$ denotes the proper time. The matter part $S_1+S_2$ contains the interaction with the gravitational perturbation and can be expressed as
    \beqn
	S_{1}+S_{2}=-\sum_{N=1,2} m_N\int_{\zeta_N} dt~\frac{1}{\gamma_N}\sqrt{1-\gamma_N^2h_{ij}(t,\bar{\bm{x}}_N (t))v_N^i v_N^j}  \ .
	\eeqn
Here, $\bar{\bm{x}}_N(t)$ denotes the trajectory of the $N$-th particle, $v_N^i=d\bar{x}_N^i/dt$ is its velocity with $v_N^2=v_N^iv_N^i$, and $\gamma_N = 1/\sqrt{1-v_N^2}$ is the Lorentz factor. After performing the Legendre transformation, the interaction Hamiltonian up to second order in $h_{ij}$ takes the form
    \beqn
    \label{eq:intHamiltonian}
	H_{\rm int}(t, \bar{\bm{x}})\eqns \sum_{N=1,2}\left[ 
    \frac{\gamma_N^3 m_N}{2}h_{ij}(t,\bar{\bm{x}}_N(t))v_N^iv_N^j
     \right. \no\\
    &&\left. \hspace{2cm}
    +\frac{3}{8}\gamma_N^5 m_N\,h_{ij}(t,\bar{\bm{x}}_N(t))
    \,h_{lm}(t,\bar{\bm{x}}_N(t))\ v_N^iv_N^jv_N^lv_N^m
    \right]\,.
	\eeqn
As discussed in the introduction, the squeezing effect is negligible. The first term represents the leading linear interaction responsible for coherent graviton emission, while the second term describes higher-order corrections that are negligible for our purposes.
We work in the interaction picture, in which the time evolution of the graviton quantum state is governed by the interaction Hamiltonian $\hat{H}_{\rm int}$. The corresponding time-evolution operator $\hat{U}(t, \bar{\bm{x}})$ is given by
    \beqn
    \label{eq:Uhat}
	\hat{U}(t, \bar{\bm{x}})\eqns \mathcal{T}\left[\exp\left(-i\int^t dt' \hat{H}_{\rm int}(t')\right)\right]
	\eeqn
where $\mathcal{T}$ denotes the time ordering operator. 
We assume a circular orbital motion with an angular frequency $\Omega$. Taking the center of mass as the origin, the tragectories of the two black holes are given by
    \beqn
    \begin{array}{lll}
    x_1= \ds\frac{m_2}{M}a \cos(\Omega t)\ , & y_1 = \ds\frac{m_2}{M}a \sin(\Omega t)\ , & z_1=0\ , \\[10pt]
    x_2= \ds\frac{m_1}{M}a \cos (\Omega t +\pi)\ , & y_2 = \ds\frac{m_1}{M}a \sin(\Omega t +\pi)\ , & z_2=0\ ,
    \end{array}
    \label{trajectories}
    \eeqn
where $M=m_1 + m_2 $ is the total mass and $a$ is the orbital separation. The time evolution operator describes the quantum state of gravitons produced by the binary system. Retaining only the linear term in $h_{ij}$ in the interaction Hamiltonian (\ref{eq:intHamiltonian}) yields an operator that generates a coherent state:
  \beqn
	\hat{U}(t, \bar{\bm{x}}_N)\eqns \exp\left[-i\frac{2}{M_{\rm p}}\sum_{N=1,2}\frac{\gamma_N^3 m_N}{2}\sum_{P=+,\times}\int^t dt'
  \right. \no\\
    &&\hspace{1.0cm}\left. \times 
    \sum_{\bm{k}}\left(\frac{e^{-i\omega_{\bm k} t'}}{\sqrt{2\omega_{\bm{k}}}}e^{P}_{ij}(\bm{k})a_{\bm{k}}^P
    +\frac{e^{i\omega_{\bm{k}}t'}}{\sqrt{2\omega_{\bm{k}}}}e^{P}_{ij}(-\bm{k})a^{P\dag}_{-\bm{k}}\right)e^{i\bm{k}\cdot\bar{\bm{x}}_N} v_N^i v_N^j\right]\ .
    \eeqn
Comparing this expression with the definition of the displacement operator,
	\beqn
    \label{eq:opecoherent}
	\hat{D}(\xi)=\prod_{\bm{k}}\prod_{P}\exp\left[\xi^{P}_{\bm k}a^{P\dag}_{\bm k}-\xi^{P*}_{\bm k}a^{P}_{\bm k}\right]\ ,
	\eeqn    
we identify the coherent parameter as
    \beqn
	\xi^{P}_{\bm{k}}\eqns -i\sum_{N=1,2}\int^t dt' \frac{\gamma_N^3 m_N}{M_{\rm p}}\frac{e^{i\omega_{\bm k} t'}}{\sqrt{2\omega_{\bm{k}}}}e^{P}_{ij}(\bm{k})v_N^i v_N^j e^{-i\bm{k}\cdot\bar{\bm{x}}_N}\ .
	\eeqn
This parameter encodes the classical orbital dynamics of the binary system into the quantum coherent state of gravitons, providing the bridge between the classical gravitational-wave signal and its quantum description.

For explicit evaluation, we choose the coordinate system such that the binary orbit lies in the $x-y$ plane plane and the wavevector ${\bm k}$ forms an angle $\theta$ with respect to the  $z$-axis, which we take to align with the position vector
$\bm{x}$. The wavevector is then parametrized as $
\bm{k}=k(\sin\theta\cos\varphi, \sin\theta\sin\varphi, \cos\theta)$. The coherent parameters then evaluate to
\beqn
    \xi^{+}_{\bm{k}}
    \eqns i\frac{1}{\sqrt{V}}\frac{\mu (a\Omega)^2}{\sqrt{2} M_{\rm p}}\int^t dt' \frac{e^{i\omega_{\bm k} t'}}{\sqrt{2\omega_{\bm{k}}}}\left(\frac{\sin^2\theta}{2}+\frac{1+\cos^2\theta}{2}\cos(2\Omega t'- 2\varphi)\right)\no\\[6pt]
    &&\hspace{3.5cm} \times\left[\gamma^3_1 \left(\frac{m_2}{M}\right)e^{-i(k_xx_1+k_yy_1)}+\gamma^3_2 \left(\frac{m_1}{M}\right)e^{-i(k_xx_2+k_yy_2)}\right]\label{eq:xikplus}\ ,\\[10pt]
    \xi^{\times}_{\bm{k}}
    \eqns i\frac{1}{\sqrt{V}}\frac{\mu (a\Omega)^2}{\sqrt{2} M_{\rm p}}\int^t dt' \frac{e^{i\omega_{\bm k} t'}}{\sqrt{2\omega_{\bm{k}}}}\cos\theta\sin(2\Omega t' - 2\varphi)\no\\[6pt]
    &&\hspace{3.5cm} \times \left[\gamma^3_1 \left(\frac{m_2}{M}\right)e^{-i(k_xx_1+k_yy_1)}+\gamma^3_2 \left(\frac{m_1}{M}\right)e^{-i(k_xx_2+k_yy_2)}\right]\label{eq:xikcloss}\ ,
    \eeqn
where we have introduced the reduced mass $\mu=(m_1 m_2) /M$. These expressions explicitly show how the polarization dependence of the emitted gravitons arises from the orbital geometry and the motion of the binary constituents. 

In summary, the binary black hole system acts as a classical source that induces a displacement on the preexisting quantum state of gravitons. Since the primordial gravitational wave background generated during inflation is already in a squeezed state, this interaction superposes a coherent displacement upon the squeezed vacuum. The role of the primordial gravitational-wave background in our setup is not to dominate the classical strain amplitude of the binary signal. Rather, it provides the squeezed component of the quantum state, while the gravitational waves emitted by the binary system provide the coherent displacement. The resulting state is therefore a coherent--squeezed state, and the nonclassicality discussed in this paper arises from the interplay between these two ingredients.

\section{Criterion for detecting gravitons}

In the previous sections, we showed that the state of gravitational waves can be described as a coherent-squeezed state. This implies that the quantum nature of primordial gravitational waves may, in principle, be probed through observation of binary black holes. In this section, we rewrite the condition~(\ref{eq:condition_cohsq}) in terms of the present-day observed frequency $f$.

The physical frequency observed today is given by
	\beqn
	2\pi f = \frac{k}{a(t_0)}\ ,
	\eeqn
where $k$ is the comoving wave number and $t_0$ is the present cosmic time. 
During inflation, the scale factor evolves as $a(\eta)=-1/H\eta$ , and thus $k\eta_1$ can be expressd as
	\beqn
	k\eta_1=2\pi f a(t_0)\eta_1=\frac{2\pi f}{H}\left(\frac{t_0}{t_{\rm eq}}\right)^{2/3}\left(\frac{t_{\rm eq}}{t_1}\right)^{1/2}
	\eeqn
where we have used the fact that the scale factor grows as $a\propto t^{1/2}$ in the radiation dominated era and $a\propto t^{2/3}$ in the matter-dominated era. Here, $t_{\rm eq}$ denotes the time of matter–radiation equality.
We introduce the cutoff frequency $f_1$ for primordial gravitational waves, corresponding to those generated at the end of inflation. 
It is defined by 
    \beqn
    \label{eq:ff1}
    k\eta_1\equiv\frac{f}{f_1}
    \eeqn
Using the redshift relation $1+z_{\rm eq}=a(t_0)/a(t_{\rm eq})=(t_0/t_{\rm eq})^{2/3}$, we can express $f_1$ as 
    \beqn
    \label{eq:cutofffreq}
    f_1=\frac{H}{2\pi}\frac{1}{1+z_{\rm eq}}\left(\frac{H_{\rm eq}}{H}\right)^{1/2}\simeq 10^{9}\sqrt{\frac{H}{10^{-4}M_{\rm p}}}~{\rm [Hz]} \ ,
    \eeqn
where we have normalized by Grand Unified Theory (GUT) energy scale, $10^{-4}M_{\rm p}$. We used the numbers $z_{\rm eq}=2.4\times 10^4$ and $1/(2H_{\rm eq})=10^{11}{\rm s}$. 
From~Eq.~(\ref{eq:Bogoliubov}), we obtain
    \beqn
    \label{eq:sinhrk}
    \sinh r_k = \frac{1}{2 k^2\eta_1^2} 
    = \frac{1}{2} \left( \frac{f_1}{f} \right)^2 \ .
    \eeqn
Considering primordial gravitational waves on super-horizon scales $r_k\gg 1$ and combining Eq.~(\ref{eq:condition_cohsq}), (\ref{eq:cutofffreq}) and (\ref{eq:sinhrk}), we obtain the condition for detecting nonclassical primordial gravitational waves as
	\beqn
	f > \left(\frac{1}{8}\right)^{\frac{1}{8}}10^{9}\left|\xi_k\right|^{-\frac{1}{4}}\sqrt{\frac{H}{10^{-4}M_{\rm p}}}~{\rm [Hz]}\ .
	\eeqn

From Eqs. (\ref{eq:xikplus}) and (\ref{eq:xikcloss}), we can estimate the order of magnitude of the coherent parameter as
   \beqn
	\left|\xi_k\right|\equiv \left|\xi_k^{+}\right|_{\rm max}=\left|\xi_k^{\times}\right|_{\rm max}=\frac{1}{\sqrt{32\pi}}\frac{1}{\sqrt{V}}\frac{\mu (a\Omega)^2}{M_{\rm p}}\frac{T}{\sqrt{f}}\left[\gamma^3_1 \left(\frac{m_2}{M}\right)+\gamma^3_2 \left(\frac{m_1}{M}\right)\right]\ .
	\eeqn
where $T$ denotes the measurement time. Here we used the relation $k=2\pi f$. 
To relate the mode spacing to the measurement time, we introduce the frequency resolution $\Delta f$ determined by $T$: 
    \beqn
    \Delta k = 2\pi \Delta f\sim \frac{2\pi}{T}\ .
    \eeqn
Since $\Delta k = \frac{2\pi}{L}$ by definition of the wavevector discretization, the effective spatial volume can be estimated as $V\sim L^3\sim T^3$. Substituting this relation, we obtain
    \beqn
	\left|\xi_k\right|=\frac{1}{\sqrt{32\pi}}\frac{\mu (a\Omega)^2}{M_{\rm p}}\frac{1}{\sqrt{Tf}}\left[\gamma^3_1 \left(\frac{m_2}{M}\right)+\gamma^3_2 \left(\frac{m_1}{M}\right)\right]
    \ .
	\eeqn
Now we evaluate the coherent parameter $\left|\xi_k\right|$ for the representative event GW150914~\cite{LIGOScientific:2016aoc}. Given the component black hole masses $36~M_{\odot}$ and $29~M_{\odot}$, the reduced mass is $\mu\sim 16~M_{\odot}$. The luminosity distance to the source is $D_L\simeq 410~{\rm Mpc}$, and the effective measurement time is $T\sim 0.2~{\rm s}$. Considering the orbital motion near the innermost stable circular orbit (ISCO), we take the orbital velocity to be $a\Omega = 1/\sqrt{6}\sim 0.41$, which corresponds to Lorentz factors $\gamma_1=1.02$ and $\gamma_2=1.03$. Substituting these values into the previous expression, we find
 \beqn
	\left|\xi_k\right|=1\times 10^{38}\left(\frac{\mu}{16~M_\odot}\right)\left(\frac{a\Omega}{0.41}\right)^2\left(\frac{0.2~{\rm s}}{T}\right)^{\frac{1}{2}}\left(\frac{68~{\rm Hz}}{f}\right)^{\frac{1}{2}},
	\eeqn
Hence, the condition for observing nonclassical primordial gravitational waves becomes
	\beqn
    \label{eq:conditionf}
	f > 0.2~{\rm Hz}~\left(\frac{\mu}{16~M_\odot}\right)^{-\frac{2}{7}}\left(\frac{a\Omega}{0.41}\right)^{-\frac{4}{7}}\left(\frac{0.2~{\rm s}}{T}\right)^{-\frac{1}{7}}\left(\frac{H}{10^{-4}M_{\rm p}}\right)^{\frac{4}{7}}\ .
	\eeqn
Equation~(\ref{eq:conditionf}) identifies the frequency range in which the coherent-squeezed graviton state can exhibit sub-Poissonian statistics. This should be understood as a condition on the underlying quantum state, not as an assessment of observational feasibility. Although this frequency band overlaps with that of ground-based interferometers such as LIGO, Virgo, and KAGRA~\cite{LIGOScientific:2025bkz}, the observable relevant to our analysis is a second-order intensity correlation rather than the strain amplitude itself. A realistic signal-to-noise analysis for such an HBT-type measurement is beyond the scope of the present paper, and we do not claim that current detectors can access this observable.

\section{Conclusion}
We have proposed a theoretical framework in which gravitational waves from binary systems can, in principle, serve as probes of nonclassical graviton states. 
Our focus is on the quantum state of gravitons, which encodes information about physical processes that occurred in the early history of the Universe. 
As an application, we examined whether the quantum nature of primordial gravitational waves can be detected through gravitational waves emitted by binary black holes. 
We showed that primordial gravitational waves generated during inflation appear as squeezed states to present-day observers. 
Furthermore, since binary systems act as classical sources producing additional displacements, the resulting gravitational-wave state can be described as a coherent–squeezed state. 
Based on this framework, we derived the condition~(\ref{eq:conditionf}) under which the graviton number statistics become sub-Poissonian—a clear indicator of nonclassicality.
In principle, this property is encoded in intensity-intensity correlations accessible through Hanbury Brown-Twiss (HBT) interferometry. Our analysis identifies the frequency range in which gravitational waves from binary black holes may carry sub-Poissonian graviton statistics. We stress, however, that the present work does not establish the observational feasibility of such an HBT-type measurement with current detectors, since a quantitative signal-to-noise analysis for intensity correlations lies beyond the scope of this paper. If such signatures were observed, they would provide compelling evidence for the quantum nature of the graviton. 

It should be noted that other mechanisms may also generate nontrivial quantum states, as mentioned in the Introduction. 
In the early Universe, various phase transitions could induce strong squeezing of gravitational modes. 
Even within the inflationary paradigm, nonstandard scenarios may produce primordial gravitational waves with distinct quantum features. 
For instance, coherent–squeezed states can arise if gauge fields are active during inflation~\cite{Maleknejad:2012fw}. 
In Refs.~\cite{Kanno:2018cuk,Kanno:2019gqw}, it was shown that the presence of classical sources during inflation allows for the detection of nonclassical primordial gravitational waves through HBT interferometry. 

Finally, we emphasize that current gravitational-wave observatories measure the strain amplitude rather than graviton number statistics directly. The nonclassical signature discussed in this paper is therefore not a modification of the mean classical waveform, but a higher-order statistical property of the radiation field, diagnosed for example by the Fano factor or the second-order correlation function. Accordingly, the present results should be interpreted as a proof-of-principle characterization of a nonclassical gravitational-wave observable, conditional on the survival of sufficient primordial quantum coherence. A quantitative study of detector sensitivity and of decoherence effects on the primordial squeezed state is left for future work. In this sense, the present analysis provides a theoretical benchmark for future quantum-sensitive gravitational-wave detection concepts aimed at probing graviton counting statistics and related higher-order observables.
Our proposed method offers a possible way to distinguish among different inflationary models and to probe unknown phenomena in the early Universe that are encoded in the quantum state of gravitons.



\section*{Acknowledgments}
S.\ K. was supported by the Japan Society for the Promotion of Science (JSPS) KAKENHI Grant Numbers JP22H01220, 24K21548 and MEXT KAKENHI Grant-in-Aid for Transformative Research Areas A “Extreme Universe” No. 24H00967.
J.\ S. was in part supported by JSPS KAKENHI Grant Numbers JP23K22491, JP24K21548, JP25H02186. A.\ T. was supported by JSPS KAKENHI Grant Number JP25KJ1912.

\printbibliography

@article{LIGOScientific:2016aoc,
    author = "Abbott, B. P. and others",
    collaboration = "LIGO Scientific, Virgo",
    title = "{Observation of Gravitational Waves from a Binary Black Hole Merger}",
    eprint = "1602.03837",
    archivePrefix = "arXiv",
    primaryClass = "gr-qc",
    reportNumber = "LIGO-P150914",
    doi = "10.1103/PhysRevLett.116.061102",
    journal = "Phys. Rev. Lett.",
    volume = "116",
    number = "6",
    pages = "061102",
    year = "2016"
}

@article{Grishchuk:1989ss,
    author = "Grishchuk, L. P. and Sidorov, Yu. V.",
    title = "{On the Quantum State of Relic Gravitons}",
    doi = "10.1088/0264-9381/6/9/002",
    journal = "Class. Quant. Grav.",
    volume = "6",
    pages = "L161--L165",
    year = "1989"
}

@article{Grishchuk:1990bj,
    author = "Grishchuk, L. P. and Sidorov, Yu. V.",
    title = "{Squeezed quantum states of relic gravitons and primordial density fluctuations}",
    doi = "10.1103/PhysRevD.42.3413",
    journal = "Phys. Rev. D",
    volume = "42",
    pages = "3413--3421",
    year = "1990"
}

@article{Dyson:2013hbl,
    author = "Dyson, Freeman",
    title = "{Is a graviton detectable?}",
    doi = "10.1142/S0217751X1330041X",
    journal = "Int. J. Mod. Phys. A",
    volume = "28",
    pages = "1330041",
    year = "2013"
}

@article{Parikh:2020nrd,
    author = "Parikh, Maulik and Wilczek, Frank and Zahariade, George",
    title = "{The Noise of Gravitons}",
    eprint = "2005.07211",
    archivePrefix = "arXiv",
    primaryClass = "hep-th",
    doi = "10.1142/S0218271820420018",
    journal = "Int. J. Mod. Phys. D",
    volume = "29",
    number = "14",
    pages = "2042001",
    year = "2020"
}

@article{Kanno:2020usf,
    author = "Kanno, Sugumi and Soda, Jiro and Tokuda, Junsei",
    title = "{Noise and decoherence induced by gravitons}",
    eprint = "2007.09838",
    archivePrefix = "arXiv",
    primaryClass = "hep-th",
    reportNumber = "OU-HET-1065, KOBE-COSMO-20-12",
    doi = "10.1103/PhysRevD.103.044017",
    journal = "Phys. Rev. D",
    volume = "103",
    number = "4",
    pages = "044017",
    year = "2021"
}

@article{Parikh:2020kfh,
    author = "Parikh, Maulik and Wilczek, Frank and Zahariade, George",
    title = "{Quantum Mechanics of Gravitational Waves}",
    eprint = "2010.08205",
    archivePrefix = "arXiv",
    primaryClass = "hep-th",
    doi = "10.1103/PhysRevLett.127.081602",
    journal = "Phys. Rev. Lett.",
    volume = "127",
    number = "8",
    pages = "081602",
    year = "2021"
}

@article{Parikh:2020fhy,
    author = "Parikh, Maulik and Wilczek, Frank and Zahariade, George",
    title = "{Signatures of the quantization of gravity at gravitational wave detectors}",
    eprint = "2010.08208",
    archivePrefix = "arXiv",
    primaryClass = "hep-th",
    doi = "10.1103/PhysRevD.104.046021",
    journal = "Phys. Rev. D",
    volume = "104",
    number = "4",
    pages = "046021",
    year = "2021"
}

@article{Kanno:2021gpt,
    author = "Kanno, Sugumi and Soda, Jiro and Tokuda, Junsei",
    title = "{Indirect detection of gravitons through quantum entanglement}",
    eprint = "2103.17053",
    archivePrefix = "arXiv",
    primaryClass = "gr-qc",
    reportNumber = "KOBE-COSMO-21-06",
    doi = "10.1103/PhysRevD.104.083516",
    journal = "Phys. Rev. D",
    volume = "104",
    number = "8",
    pages = "083516",
    year = "2021"
}

@article{Manikandan:2025dea,
    author = "Manikandan, Sreenath K. and Wilczek, Frank",
    title = "{Squeezed Quasinormal Modes from Nonlinear Gravitational Effects}",
    eprint = "2508.03380",
    archivePrefix = "arXiv",
    primaryClass = "gr-qc",
    month = "8",
    year = "2025"
}

@article{Guerreiro:2025sge,
    author = "Guerreiro, Thiago",
    title = "{Entanglement and squeezing of gravitational waves}",
    eprint = "2501.17043",
    archivePrefix = "arXiv",
    primaryClass = "gr-qc",
    month = "1",
    year = "2025"
}

@article{Dorlis:2025zzz,
    author = "Dorlis, Panagiotis and Mavromatos, Nick E. and Sarkar, Sarben and Vlachos, Sotirios-Neilos",
    title = "{Superradiant Axionic Black-Hole Clouds as Seeds for Graviton Squeezing}",
    eprint = "2507.01689",
    archivePrefix = "arXiv",
    primaryClass = "gr-qc",
    reportNumber = "KCL-PH-TH/2025-15",
    month = "7",
    year = "2025"
}

@article{Dorlis:2025amf,
    author = "Dorlis, Pangiotis and Mavromatos, Nick E. and Sarkar, Sarben and Vlachos, Sotirios-Neilos",
    title = "{Squeezed gravitons from superradiant axion fields around rotating black holes}",
    eprint = "2507.23475",
    archivePrefix = "arXiv",
    primaryClass = "gr-qc",
    reportNumber = "KCL-PH-TH/2025-26",
    month = "7",
    year = "2025"
}

@article{HanburyBrown:1956bqd,
    author = "Hanbury Brown, R. and Twiss, R. Q.",
    title = "{A Test of a new type of stellar interferometer on Sirius}",
    doi = "10.1038/1781046a0",
    journal = "Nature",
    volume = "178",
    pages = "1046--1048",
    year = "1956"
}

@article{Brown:1956zza,
    author = "Brown, R. Hanbury and Twiss, R. Q.",
    title = "{Correlation between Photons in two Coherent Beams of Light}",
    doi = "10.1038/177027a0",
    journal = "Nature",
    volume = "177",
    pages = "27--29",
    year = "1956"
}

@article{Kanno:2018cuk,
    author = "Kanno, Sugumi and Soda, Jiro",
    title = "{Detecting nonclassical primordial gravitational waves with Hanbury-Brown{\textendash}Twiss interferometry}",
    eprint = "1810.07604",
    archivePrefix = "arXiv",
    primaryClass = "hep-th",
    reportNumber = "OU-HET-980, KOBE-COSMO-18-09",
    doi = "10.1103/PhysRevD.99.084010",
    journal = "Phys. Rev. D",
    volume = "99",
    number = "8",
    pages = "084010",
    year = "2019"
}

@article{Hertzberg:2021rbl,
    author = "Hertzberg, Mark P. and Litterer, Jacob A.",
    title = "{Bound on quantum fluctuations in gravitational waves from LIGO-Virgo}",
    eprint = "2112.12159",
    archivePrefix = "arXiv",
    primaryClass = "gr-qc",
    doi = "10.1088/1475-7516/2023/03/009",
    journal = "JCAP",
    volume = "03",
    pages = "009",
    year = "2023"
}

@article{Kanno:2025how,
    author = "Kanno, Sugumi and Soda, Jiro and Taniguchi, Akira",
    title = "{Quantum nature of gravitational waves from binary black holes}",
    eprint = "2508.17947",
    archivePrefix = "arXiv",
    primaryClass = "gr-qc",
    reportNumber = "YITP-25-129, KOBE-COSMO-25-16",
    month = "8",
    year = "2025"
}

@article{Giovannini:2010xg,
    author = "Giovannini, Massimo",
    title = "{Hanbury Brown-Twiss interferometry and second-order correlations of inflaton quanta}",
    eprint = "1011.1673",
    archivePrefix = "arXiv",
    primaryClass = "astro-ph.CO",
    reportNumber = "CERN-PH-TH-2010-242",
    doi = "10.1103/PhysRevD.83.023515",
    journal = "Phys. Rev. D",
    volume = "83",
    pages = "023515",
    year = "2011"
}

@article{Giovannini:2016esa,
    author = "Giovannini, Massimo",
    title = "{Glauber theory and the quantum coherence of curvature inhomogeneities}",
    eprint = "1608.05843",
    archivePrefix = "arXiv",
    primaryClass = "hep-th",
    reportNumber = "CERN-TH-2016-181",
    doi = "10.1088/1361-6382/aa52d9",
    journal = "Class. Quant. Grav.",
    volume = "34",
    number = "3",
    pages = "035019",
    year = "2017"
}

@article{Giovannini:2017uty,
    author = "Giovannini, Massimo",
    title = "{Quantum coherence of cosmological perturbations}",
    eprint = "1709.00914",
    archivePrefix = "arXiv",
    primaryClass = "gr-qc",
    doi = "10.1142/S0217732317501917",
    journal = "Mod. Phys. Lett. A",
    volume = "32",
    number = "35",
    pages = "1750191",
    year = "2017"
}

@article{Kanno:2019gqw,
    author = "Kanno, Sugumi",
    title = "{Nonclassical primordial gravitational waves from the initial entangled state}",
    eprint = "1905.06800",
    archivePrefix = "arXiv",
    primaryClass = "hep-th",
    reportNumber = "OU-HET-1017",
    doi = "10.1103/PhysRevD.100.123536",
    journal = "Phys. Rev. D",
    volume = "100",
    number = "12",
    pages = "123536",
    year = "2019"
}

@article{Kanno:2024gjt,
    author = "Kanno, Sugumi and Matsui, Hiroki and Mukohyama, Shinji",
    title = "{Hanbury-Brown-Twiss interferometry and quantum nature of primordial gravitational waves in Ho{\v{r}}ava-Lifshitz gravity}",
    eprint = "2412.19514",
    archivePrefix = "arXiv",
    primaryClass = "gr-qc",
    reportNumber = "YITP-24-160, IPMU24-0044",
    doi = "10.1103/PhysRevD.111.104077",
    journal = "Phys. Rev. D",
    volume = "111",
    number = "10",
    pages = "104077",
    year = "2025"
}

@article{Kawamura:2011zz,
    author = "Kawamura, Seiji and others",
    editor = "Buchman, Sasha and Sun, Ke-Xun",
    title = "{The Japanese space gravitational wave antenna: DECIGO}",
    doi = "10.1088/0264-9381/28/9/094011",
    journal = "Class. Quant. Grav.",
    volume = "28",
    pages = "094011",
    year = "2011"
}

@article{LISACosmologyWorkingGroup:2022jok,
    author = "Auclair, Pierre and others",
    collaboration = "LISA Cosmology Working Group",
    title = "{Cosmology with the Laser Interferometer Space Antenna}",
    eprint = "2204.05434",
    archivePrefix = "arXiv",
    primaryClass = "astro-ph.CO",
    reportNumber = "LISA CosWG-22-03, FERMILAB-PUB-22-349-SCD",
    doi = "10.1007/s41114-023-00045-2",
    journal = "Living Rev. Rel.",
    volume = "26",
    number = "1",
    pages = "5",
    year = "2023"
}

@book{agarwal2013quantum,
  title={Quantum optics},
  author={Agarwal, Girish S},
  year={2013},
  publisher={Cambridge University Press}
}

@article{Maleknejad:2012fw,
    author = "Maleknejad, A. and Sheikh-Jabbari, M. M. and Soda, J.",
    title = "{Gauge Fields and Inflation}",
    eprint = "1212.2921",
    archivePrefix = "arXiv",
    primaryClass = "hep-th",
    reportNumber = "IPM-P-2012-049, KUNS-2430",
    doi = "10.1016/j.physrep.2013.03.003",
    journal = "Phys. Rept.",
    volume = "528",
    pages = "161--261",
    year = "2013"
}

@article{Ito:2019wcb,
    author = "Ito, Asuka and Ikeda, Tomonori and Miuchi, Kentaro and Soda, Jiro",
    title = "{Probing GHz gravitational waves with graviton\textendash{}magnon resonance}",
    eprint = "1903.04843",
    archivePrefix = "arXiv",
    primaryClass = "gr-qc",
    reportNumber = "KOBE-COSMO-19-01",
    doi = "10.1140/epjc/s10052-020-7735-y",
    journal = "Eur. Phys. J. C",
    volume = "80",
    number = "3",
    pages = "179",
    year = "2020"
}

@article{Ejlli:2019bqj,
    author = "Ejlli, Aldo and Ejlli, Damian and Cruise, Adrian Mike and Pisano, Giampaolo and Grote, Hartmut",
    title = "{Upper limits on the amplitude of ultra-high-frequency gravitational waves from graviton to photon conversion}",
    eprint = "1908.00232",
    archivePrefix = "arXiv",
    primaryClass = "gr-qc",
    doi = "10.1140/epjc/s10052-019-7542-5",
    journal = "Eur. Phys. J. C",
    volume = "79",
    number = "12",
    pages = "1032",
    year = "2019"
}

@article{Ito:2020wxi,
    author = "Ito, Asuka and Soda, Jiro",
    title = "{A formalism for magnon gravitational wave detectors}",
    eprint = "2004.04646",
    archivePrefix = "arXiv",
    primaryClass = "gr-qc",
    reportNumber = "KOBE-COSMO-20-07",
    doi = "10.1140/epjc/s10052-020-8092-6",
    journal = "Eur. Phys. J. C",
    volume = "80",
    number = "6",
    pages = "545",
    year = "2020"
}

@article{Berlin:2021txa,
    author = {Berlin, Asher and Blas, Diego and Tito D'Agnolo, Raffaele and Ellis, Sebastian A. R. and Harnik, Roni and Kahn, Yonatan and Sch\"utte-Engel, Jan},
    title = "{Detecting high-frequency gravitational waves with microwave cavities}",
    eprint = "2112.11465",
    archivePrefix = "arXiv",
    primaryClass = "hep-ph",
    reportNumber = "FERMILAB-PUB-21-724-SQMS-T",
    doi = "10.1103/PhysRevD.105.116011",
    journal = "Phys. Rev. D",
    volume = "105",
    number = "11",
    pages = "116011",
    year = "2022"
}

@article{Domcke:2022rgu,
    author = "Domcke, Valerie and Garcia-Cely, Camilo and Rodd, Nicholas L.",
    title = "{Novel Search for High-Frequency Gravitational Waves with Low-Mass Axion Haloscopes}",
    eprint = "2202.00695",
    archivePrefix = "arXiv",
    primaryClass = "hep-ph",
    reportNumber = "DESY-22-017, CERN-TH-2022-010",
    doi = "10.1103/PhysRevLett.129.041101",
    journal = "Phys. Rev. Lett.",
    volume = "129",
    number = "4",
    pages = "041101",
    year = "2022"
}

@article{Ito:2022rxn,
    author = "Ito, Asuka and Soda, Jiro",
    title = "{Exploring high-frequency gravitational waves with magnons}",
    eprint = "2212.04094",
    archivePrefix = "arXiv",
    primaryClass = "gr-qc",
    reportNumber = "KEK-QUP-2022-0018, KEK-TH-2477, KEK-Cosmo-0304, KOBE-COSMO-22-19",
    doi = "10.1140/epjc/s10052-023-11876-2",
    journal = "Eur. Phys. J. C",
    volume = "83",
    number = "8",
    pages = "766",
    year = "2023"
}

@article{Tobar:2024bjr,
    author = "Tobar, Germain and Pikovski, Igor and Tobar, Michael Edmund",
    title = "{Detecting kHz gravitons from a neutron star merger with a multi-mode resonant mass detector}",
    eprint = "2406.16898",
    archivePrefix = "arXiv",
    primaryClass = "astro-ph.IM",
    doi = "10.1088/1361-6382/adae4a",
    journal = "Class. Quant. Grav.",
    volume = "42",
    number = "5",
    pages = "055017",
    year = "2025"
}

@article{Kanno:2023whr,
    author = "Kanno, Sugumi and Soda, Jiro and Taniguchi, Akira",
    title = "{Search for high-frequency gravitational waves with Rydberg atoms}",
    eprint = "2311.03890",
    archivePrefix = "arXiv",
    primaryClass = "gr-qc",
    reportNumber = "KOBE-COSMO-23-10",
    doi = "10.1140/epjc/s10052-024-13736-z",
    journal = "Eur. Phys. J. C",
    volume = "85",
    number = "1",
    pages = "31",
    year = "2025"
}

@article{Ikeda:2025uae,
    author = "Ikeda, Taiki and Kaku, Youka and Kanno, Sugumi and Soda, Jiro",
    title = "{Toward graviton detection via photon-graviton quantum state conversion}",
    eprint = "2507.01609",
    archivePrefix = "arXiv",
    primaryClass = "quant-ph",
    reportNumber = "KOBE-COSMO-25-12",
    month = "7",
    year = "2025"
}

@article{Ikeda:2025qac,
    author = "Ikeda, Taiki and Kanno, Sugumi and Soda, Jiro",
    title = "{Enhancing photon-axion conversion probability with squeezed coherent states}",
    eprint = "2506.14354",
    archivePrefix = "arXiv",
    primaryClass = "quant-ph",
    reportNumber = "YITP-25-105, KOBE-COSMO-25-11",
    month = "6",
    year = "2025"
}

@article{LIGOScientific:2025bkz,
    author = "Abac, A. G. and others",
    collaboration = "LIGO Scientific, VIRGO, KAGRA",
    title = "{Directional Search for Persistent Gravitational Waves: Results from the First Part of LIGO-Virgo-KAGRA's Fourth Observing Run}",
    eprint = "2510.17487",
    archivePrefix = "arXiv",
    primaryClass = "gr-qc",
    reportNumber = "LIGO-P250038",
    month = "10",
    year = "2025"
}
\end{document}